\documentclass[baaa]{baaa}

\usepackage[pdftex]{hyperref}
\usepackage{subfigure}
\usepackage{natbib}
\usepackage{helvet,soul}
\usepackage[font=small]{caption}

\contriblanguage{0}

\contribtype{1}

\thematicarea{7}

\title{Estrellas híbridas magnetizadas en el contexto de astronomía multimensajera}

\titlerunning{Estrellas híbridas magnetizadas}

\author{
M. Mariani\inst{1, 2},
D. Curin\inst{1},
M.G. Orsaria\inst{1, 2} \&
I.F. Ranea-Sandoval\inst{1, 2}
}

\authorrunning{Mariani et al.}

\contact{mmariani@fcaglp.unlp.edu.ar}

\institute{
Grupo de Gravitaci\'on, Astrof\'isica y Cosmolog\'ia, Facultad de Ciencias Astronómicas y Geofísicas, UNLP, Argentina
\and
Consejo Nacional de Investigaciones Cient\'ificas y T\'ecnicas, Argentina
}

\resumen{Las recientes detecciones de LIGO/Virgo y NICER han puesto fuertes restricciones sobre las propiedades de las estrellas de neutrones. En este trabajo, estudiamos estrellas de neutrones modelándolas como estrellas híbridas, objetos compactos con un núcleo de materia de quarks rodeado por capas de materia hadrónica. Además, consideramos la presencia de intensos campos magnéticos, motivada por la detección de magnetares como el recientemente descubierto Swift J1818.0-1607. Incorporamos los efectos del momento magnético anómalo de las partículas constituyentes a la ecuación de estado de la materia densa y analizamos las implicancias de diferentes transiciones de fase hadrón-quark sobre la estabilidad dinámica de dichos objetos compactos. Los resultados de este estudio muestran que las restricciones sobre la masa, radio y deformabilidad de marea, impuestas por las observaciones de púlsares masivos y ondas gravitacionales, pueden ser satisfechas en el marco de nuestro modelo.}

\abstract{The most recent detections of LIGO/Virgo and NICER have placed strong constraints on neutron stars' properties. In this work, we study neutron stars modeling them as hybrid stars, compact objects with a quark matter core surrounded by layers of hadronic matter. In addition, we consider the presence of strong magnetic fields, motivated by magnetars detection, like the recently discovered Swift J1818.0-1607. We incorporate the effects of the anomalous magnetic moment of the constituent particles into the equation of state of dense matter and analyze the implications of different hadron-quark phase transitions on the dynamic stability of these compact objects. This study shows that the constraints on the mass, radius, and tidal deformability, imposed by observations of massive pulsars and gravitational waves can be satisfied within the framework of our model.}


\begin{document}

\maketitle

\section{Introducción}\label{S_intro}

Los resultados de la última detección conjunta en \mbox{rayos-X} de NICER y XMM-Newton del púlsar J0740+6620 \citep{Riley:2021anv,Miller:2021tro} (cuya masa fue establecida previamente en \cite{Cromartie:2020rsd,Fonseca:2021rma}) han entrado en conflicto con los provenientes de la astronomía multimensajera (ondas gravitacionales con contraparte electromagnética) presentados por LIGO/Virgo para el evento GW170817 \citep{Abbot:2018gmo}. Estas observaciones, junto con otras, como la de los púlsares PSR J0348+0432 \citep{Antoniadis:2013amp}, PSR J1614-2230 \citep{Demorest:2010ats,Arzoumanian:2018tny}, PSR J0030+0451 \citep{Riley:2019anv,Miller:2019pma} y la del evento GW190425 \citep{Abbot:2020goo}, establecen un conjunto de restricciones que desafían fuertemente a los actuales modelos teóricos que describen la composición de los núcleos de estrellas de neutrones (ENs).

En este trabajo, modelamos ENs como \emph{estrellas híbridas} (EHs), objetos compactos con un núcleo de materia de quarks rodeado por capas de materia hadrónica. Además, consideramos la presencia de intensos campos magnéticos (CM), motivada por la detección de \emph{magnetares}, como el recientemente descubierto Swift J1818.0-1607 \citep{Blumer:2020coo}.

\section{Ecuación de estado híbrida}
   
\begin{figure}[!t]
    \centering
    \includegraphics[width=0.95\columnwidth]{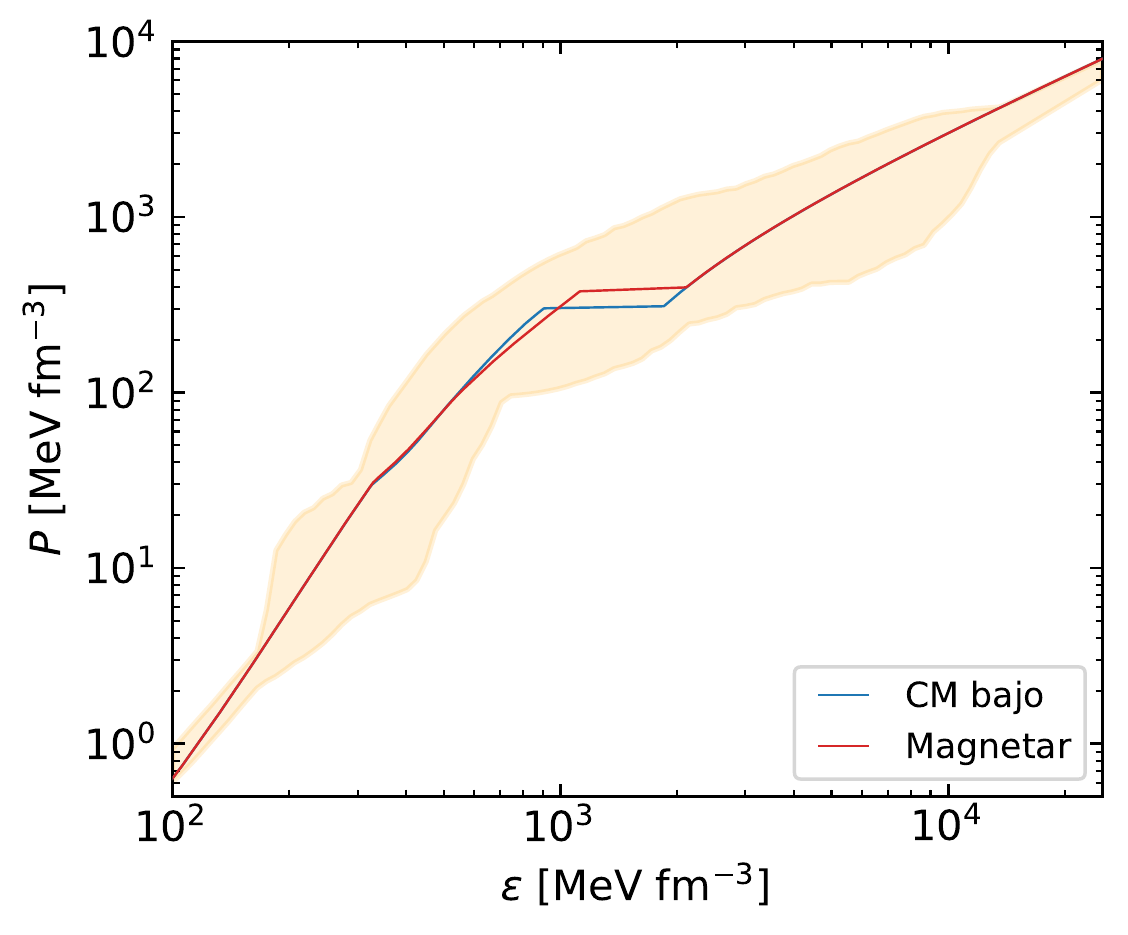}
    \caption{EdEs híbridas para ambos casos del CM. La región a presión constante señala la transición de fase hadrón-quark abrupta; densidades de energía menores a esta región corresponden a la fase de hadrones, y densidades mayores, a la fase de quarks. La región coloreada naranja corresponde a las cotas mostradas en \cite{Annala:2020efq}.}
    \label{fig:eos}%
\end{figure}

Construimos las EHs a través de una \emph{ecuación de estado} (EdE), que describe la relación entre la densidad de energía, $\epsilon$, y la presión, $P$, de la materia. A densidades bajas, consideramos la presencia de materia hadrónica -incluyendo las partículas del octeto bariónico y los bariones $\Delta$- a través del modelo Relativista de Campo Medio, teniendo en cuenta la parametrización SW4L \citep{Spinella:2018dab, Malfatti:2020onm}. Para densidades altas, consideramos la presencia de materia de quarks extraña -quarks $u$, $d$, $s$- utilizando el modelo de Campo Correlacionador (MCC) \citep{Simonov:2007dtf, Simonov:2007vpt}. Entre ambas fases, asumimos que se produce una transición de fase abrupta a presión constante, conocida como \emph{transición de Maxwell} (ver Fig.~\ref{fig:eos}).

Incluimos el CM en el interior de la estrella mediante un campo parametrizado a través de una función exponencial \citep{Dexheimer:2012hsi}. El efecto de dicho CM se tiene en cuenta en la aparición de los \emph{niveles de Landau} de las partículas cargadas y en la inclusión del \emph{momento magnético anómalo} (MMA) de las partículas eléctricamente neutras \citep{Ferrer:2015iot}.

Consideramos dos casos de estudio (ver Tabla~\ref{tabla}), tomando un solo conjunto de parámetros del MCC, dado por $V_1 = 60$~MeV, $G_2 = 0.012$~GeV$^4$. Esta elección permite analizar y comparar los resultados obtenidos para \emph{púlsares} convencionales (CM bajo) y magnetares.

Las EdEs híbridas se muestran en la Fig.~\ref{fig:eos}. A medida que el CM aumenta, la transición de fase hadron-quark ocurre a presión y densidad de energía mayores y el salto en densidad de energía se acorta. Ambos casos satisfacen las cotas -de experimentos nucleares y de la QCD perturbativa- presentadas por \cite{Annala:2020efq}.

\begin{table}[!h]
\centering
\begin{tabular}{cccc}
\hline\hline\noalign{\smallskip}
Caso & & B$_{\textrm{sup}}$~[Gauss] & B$_{\textrm{cen}}$~[Gauss] \\
\hline\noalign{\smallskip}
CM bajo & & $1.0 \times 10^{12}$ & $1.0 \times 10^{15}$  \\
Magnetar & & $1.0 \times 10^{15}$ & $3.0 \times 10^{18}$ \\
\hline
\end{tabular}
\caption{Casos de estudio seleccionados según diferentes intensidades del CM. Para el caso de CM bajo, los efectos de los niveles de Landau y del MMA son despreciables.}
\label{tabla}
\end{table}

\begin{figure}[!t]
    \centering
    \includegraphics[width=1\columnwidth]{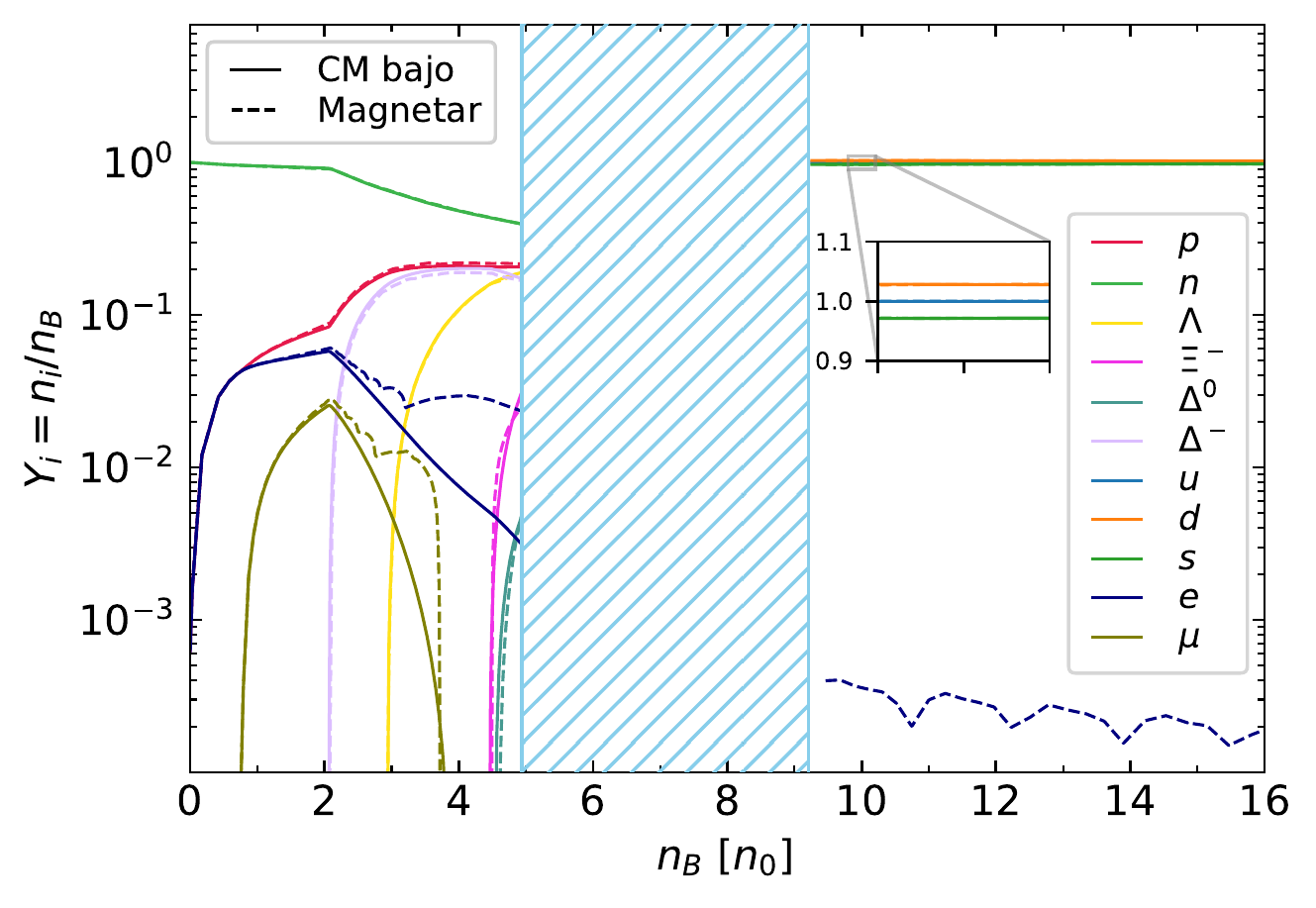}
    \caption{Abundancia de partículas en función de la densidad numérica de bariones, en unidades de la densidad de saturación nuclear, $n_0 \approx 0.16$~fm$^{-3}$, para ambos casos del CM. La región rayada celeste muestra el salto discontinuo en la densidad, producto de la transición de fase abrupta. La curvas de los quarks $u$, $d$ y $s$ se encuentran notoriamente superpuestas, como se aprecia en el recuadro aumentado; la abundancia de electrones en la fase de quarks es despreciable para el caso de CM bajo.} 
    \label{fig:pop}%
\end{figure}

En la Fig.~\ref{fig:pop} se muestra la abundancia de partículas. Se observa que la magnitud del CM no modifica la especie de partículas presentes en cada fase, aunque sí tiene efectos sobre la abundancia de cada especie. En particular, la población leptónica ($e$, $\mu$) es la más afectada por la presencia del CM, disminuyendo rápidamente para CM bajos, hasta hacerse despreciable en la fase de quarks. Además, en el caso de CM altos, las abundancias presentan oscilaciones, efecto característico de la cuantización de Landau.
    
\section{Estructura de estrellas híbridas}

 Una vez construida la EdE híbrida, modelamos las EHs integrando las ecuaciones Tolman-Oppenheimer-Volkoff de equilibrio hidrostático relativista, para obtener las configuraciones de equilibrio. Trabajamos en el marco de la hipótesis de CM caótico \citep{Mariani:2019mhs}, que permite promediar las presiones paralela y perpendicular, propias de la anisotropía local del CM, y considerar la hipótesis de simetría esférica.
 
\begin{figure}[!t]
    \centering
    \includegraphics[width=0.95\columnwidth]{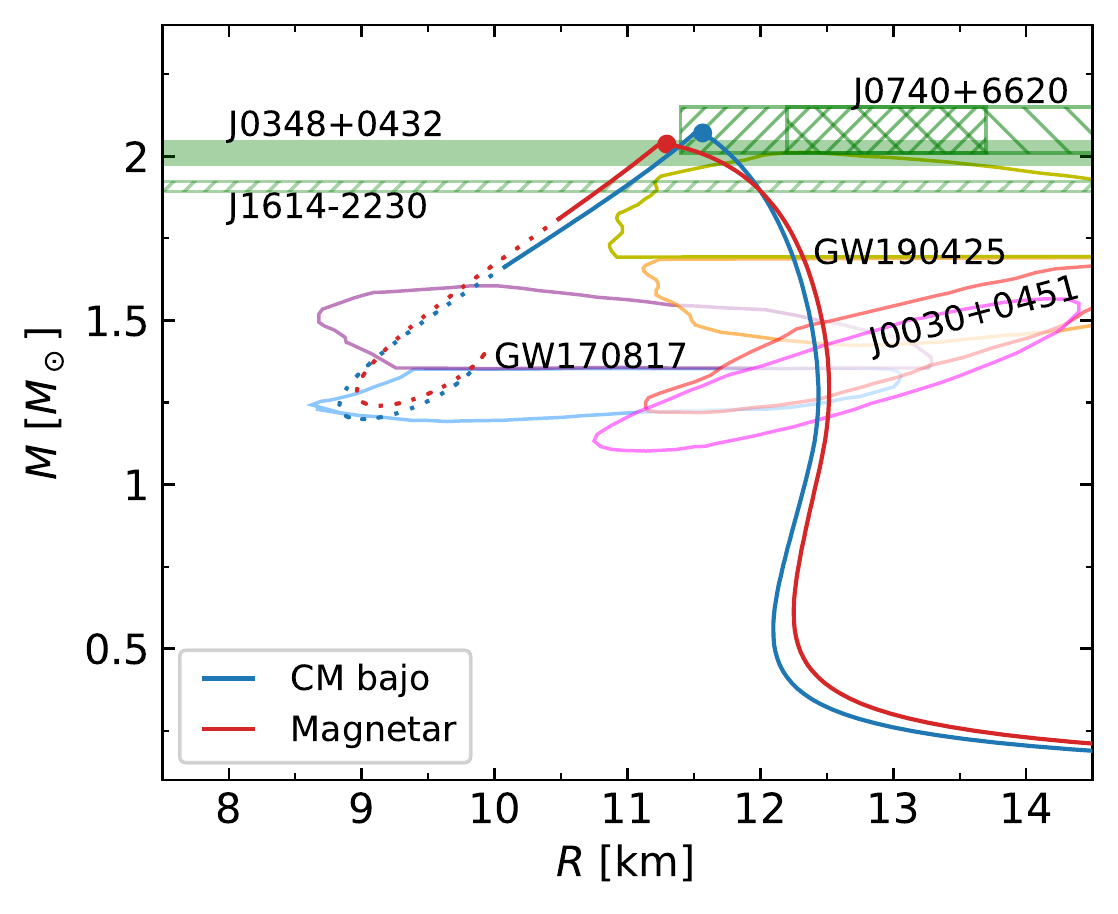}
    \caption{Plano masa-radio para ambos casos del CM. Para conversiones lentas, las soluciones estables corresponden a las curvas continuas; para conversiones rápidas, las configuraciones se desestabilizan luego del máximo, hacia radios menores . El punto circular sobre cada curva indica la aparición del núcleo de quarks. Las distintas regiones coloreadas muestran las restricciones observacionales actuales.}
    \label{fig:mraio}%
\end{figure}

En la Fig.~\ref{fig:mraio} está representada la relación masa-radio para las familias de estrellas de los casos seleccionados. En este trabajo, al considerar el escenario de conversión hadron-quark lenta, hemos encontrando largas ramas de estabilidad extendida. Cabe destacar que, en caso de considerar conversión rápida, la aparición de materia de quarks en el núcleo de estos objetos compactos desestabiliza las estrellas, dejando solamente estable la rama puramente hadrónica. De esta manera, la presencia de EHs es no despreciable solo en el caso de conversiones lentas. 

Para los casos estudiados, el aumento del CM genera que la masa máxima de las familias de estrellas disminuya (ver Tabla~\ref{tabla2}). La presencia de CM no solo afecta la masa máxima, sino que además, en el caso de conversiones lentas, las soluciones para CM bajos presentan ramas extendidas más largas, llegando hasta radios y masas menores (ver Tabla~\ref{tabla2}). Finalmente, vemos que nuestro modelo satisface las actuales restricciones observacionales para el caso de CM bajo, mientras que el caso magnetar resulta fuera de las últimas restricciones impuestas por el púlsar J0740+6620. Este escenario no resulta problemático, dado que para este púlsar particular no se estiman CM intensos.

\begin{figure}[!t]
    \centering
    \includegraphics[width=0.95\columnwidth]{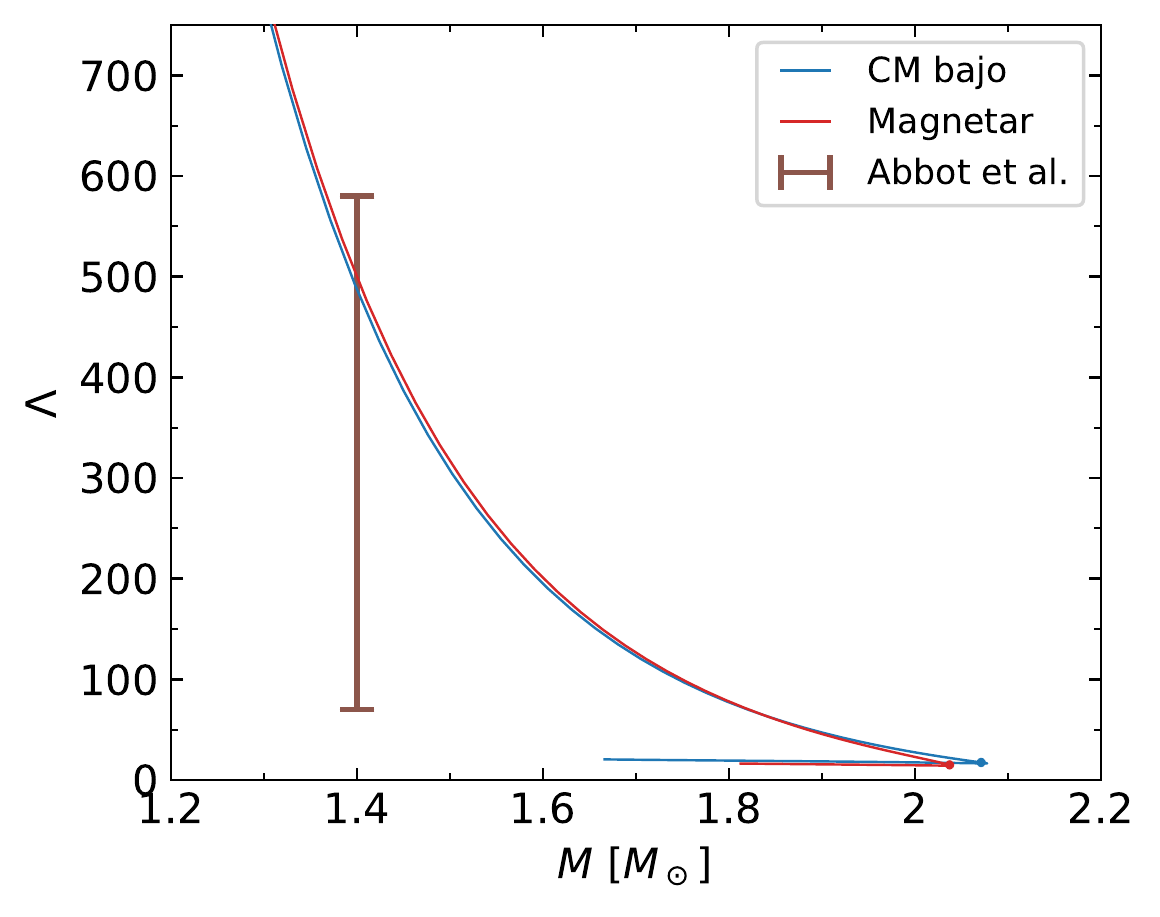}
    \caption{Plano de deformabilidad de marea adimensional, $\Lambda$-masa para ambos casos del CM. Para las conversiones lentas, solo se muestran las configuraciones de equilibrio; para el caso de conversiones rápidas, la rama estable se mantiene, desde valores de $\Lambda$ mayores ($\Lambda \sim 700$), hasta la configuración de masa máxima ($\Lambda \sim 0$). El segmento vertical indica la restricción obtenida a partir del evento GW170817.}
    \label{fig:tidal}%
\end{figure}

En la Fig.~\ref{fig:tidal} está representada la relación entre la deformabilidad de marea adimensional, $\Lambda$, y la masa para los casos de estudio seleccionados. Esta relación resulta de particular relevancia pues, a partir del evento GW170817, se ha establecido una cota para una EN con $ M = 1.4~M_\odot$, $70 < \Lambda_{1.4M_\odot} < 580$. En nuestro modelo, esta restricción se satisface a través de la rama de estabilidad tradicional -estable en ambos escenarios, rápido y lento-, que corresponde a estrellas puramente hadrónicas.

\begin{table}[!h]
\centering
\begin{tabular}{cccccc}
\hline\hline\noalign{\smallskip}
Caso & & $M_{\textrm{max}}$ & $R_{\textrm{max}}$ & $M_{\textrm{term}}$ & $R_{\textrm{term}}$ \\
\hline\noalign{\smallskip}
CM bajo & & $2.08$ & $11.52$ & $1.67$  & $10.09$  \\   
Magnetar & & $2.04$ & $11.26$ & $1.81$ & $10.49$ \\   
\hline
\end{tabular}
\caption{Valores característicos para los casos de estudio. Las masas se expresan en unidades de masas solares, $M_\odot$, y los radios en km. La etiqueta \emph{term} representa, para conversiones lentas, la última estrella estable de la rama extendida. }
\label{tabla2}
\end{table}

\section{Discusión y conclusiones}

En los últimos años han surgido variadas restricciones vinculadas a las las ENs, provenientes de experimentos de física nuclear, cálculos de la QCD, observaciones de púlsares y fusión de ENs. La discrepancia que surge entre los resultados provenientes de estas restricciones y los modelos teóricos actuales que describen las ENs aún no está resuelta.

En particular, en nuestro estudio centrado en EHs con CMs, logramos satisfacer las restricciones observacionales actuales y, además, al momento de comparar los diferentes casos de CM, evidenciamos efectos medibles sobre la estructura y estabilidad de las EHs. Queda, como trabajo a futuro, realizar una exploración sistemática del espacio de parámetros de nuestro modelo y el calculo y análisis de nuevas cantidades observables, como las frecuencias de los modos de oscilación no-radiales, relacionados con la emisión de ondas gravitacionales.

Nuestros resultados apuntan a que el estudio pormenorizado de posibles nuevos efectos incluidos en las EdEs, junto con futuras observaciones, contribuya a resolver las discrepancias actuales entre los experimentos y observaciones con la teoría.

\begin{acknowledgement}
M.M. es becario posdoctoral de CONICET, D.C. becaria doctoral de la UNLP. El trabajo de M.M., D.C., M.O. e I.F.R-S. está parcialmente financiado por el CONICET y la UNLP (Argentina) a través de los subsidios PIP-0714, G157 y G007. I.F.R-S. también se encuentra financiado parcialmente por el subsidio PICT 2019-0366 de la ANPCyT, Argentina.
\end{acknowledgement}

%%%%%%%%%%%%%%%%%%%%%%%%%%%%%%%%%%%%%%%%%%%%%%%%%%%%%%%%%%%%%%%%%%%%%%%%%%%%%%
%  ******************* Bibliografía / Bibliography ************************  %
%                                                                            %
%  -Ver en la sección 3 "Bibliografía" para mas información.                 %
%  -Debe usarse BIBTEX.                                                      %
%  -NO MODIFIQUE las líneas de la bibliografía, salvo el nombre del archivo  %
%   BIBTEX con la lista de citas (sin la extensión .BIB).                    %
%                                                                            %
%  -BIBTEX must be used.                                                     %
%  -Please DO NOT modify the following lines, except the name of the BIBTEX  %
%  file (whithout the .BIB extension).                                       %
%%%%%%%%%%%%%%%%%%%%%%%%%%%%%%%%%%%%%%%%%%%%%%%%%%%%%%%%%%%%%%%%%%%%%%%%%%%%%% 

\bibliographystyle{baaa}
\small
\bibliography{bibliografia}

\begin{thebibliography}{20}
\providecommand{\natexlab}[1]{#1}

\bibitem[{Abbott et~al.(2018)}]{Abbot:2018gmo}
Abbott B.P., et~al., 2018, Phys. Rev. Lett., 121, 161101

\bibitem[{{Abbott} et~al.(2020)}]{Abbot:2020goo}
{Abbott} B.P., et~al., 2020, ApJL, 892, L3

\bibitem[{{Annala} et~al.(2020)}]{Annala:2020efq}
{Annala} E., et~al., 2020, Nature Physics, 16, 907

\bibitem[{{Antoniadis} et~al.(2013)}]{Antoniadis:2013amp}
{Antoniadis} J., et~al., 2013, Science, 340, 448

\bibitem[{Arzoumanian et~al.(2018)}]{Arzoumanian:2018tny}
Arzoumanian Z., et~al., 2018, ApJS, 235, 37

\bibitem[{{Blumer} \& {Safi-Harb}(2020)}]{Blumer:2020coo}
{Blumer} H., {Safi-Harb} S., 2020, ApJL, 904, L19

\bibitem[{{Cromartie} et~al.(2020)}]{Cromartie:2020rsd}
{Cromartie} H.T., et~al., 2020, Nature Astronomy, 4, 72

\bibitem[{{Demorest} et~al.(2010)}]{Demorest:2010ats}
{Demorest} P.B., et~al., 2010, Nature, 467, 1081

\bibitem[{{Dexheimer} et~al.(2012)}]{Dexheimer:2012hsi}
{Dexheimer} V., et~al., 2012, European Phys. J. A, 48, 189

\bibitem[{Ferrer et~al.(2015)}]{Ferrer:2015iot}
Ferrer E.J., et~al., 2015, Phys. Rev. D, 91, 085041

\bibitem[{{Fonseca} et~al.(2021)}]{Fonseca:2021rma}
{Fonseca} E., et~al., 2021, ApJL, 915, L12

\bibitem[{Malfatti et~al.(2020)}]{Malfatti:2020onm}
Malfatti G., et~al., 2020, Phys. Rev. D, 102, 063008

\bibitem[{{Mariani} et~al.(2019)}]{Mariani:2019mhs}
{Mariani} M., et~al., 2019, MNRAS, 489, 4261

\bibitem[{{Miller} et~al.(2019)}]{Miller:2019pma}
{Miller} M.C., et~al., 2019, ApJL, 887, L24

\bibitem[{{Miller} et~al.(2021)}]{Miller:2021tro}
{Miller} M.C., et~al., 2021, arXiv:, arXiv:2105.06979

\bibitem[{{Riley} et~al.(2019)}]{Riley:2019anv}
{Riley} T.E., et~al., 2019, ApJL, 887, L21

\bibitem[{{Riley} et~al.(2021)}]{Riley:2021anv}
{Riley} T.E., et~al., 2021, arXiv:, arXiv:2105.06980

\bibitem[{{Simonov} \& {Trusov}(2007{\natexlab{a}})}]{Simonov:2007dtf}
{Simonov} Y.A., {Trusov} M.A., 2007{\natexlab{a}}, JETP Letters, 85, 598

\bibitem[{{Simonov} \& {Trusov}(2007{\natexlab{b}})}]{Simonov:2007vpt}
{Simonov} Y.A., {Trusov} M.A., 2007{\natexlab{b}}, Phys. Letters B, 650, 36

\bibitem[{Spinella \& Weber(2019)}]{Spinella:2018dab}
Spinella W.M., Weber F., 2019, Astron. Nachr., 340, 145

\end{thebibliography}
 
\end{document}